\documentclass[english,prl,reprint,twocolumn]{revtex4}
\usepackage{ae,aecompl}

\usepackage[T1]{fontenc}
\usepackage[latin9]{inputenc}
\usepackage{amsbsy}
\usepackage{amssymb}
\usepackage{graphicx}
\usepackage{esint}

\makeatletter

\@ifundefined{textcolor}{}
{%
 \definecolor{BLACK}{gray}{0}
 \definecolor{WHITE}{gray}{1}
 \definecolor{RED}{rgb}{1,0,0}
 \definecolor{GREEN}{rgb}{0,1,0}
 \definecolor{BLUE}{rgb}{0,0,1}
 \definecolor{CYAN}{cmyk}{1,0,0,0}
 \definecolor{MAGENTA}{cmyk}{0,1,0,0}
 \definecolor{YELLOW}{cmyk}{0,0,1,0}
 }

\makeatother

\usepackage{babel}
\begin{document}

\title{A witness for coherent electronic oscillations in ultrafast spectroscopy}

\author{Joel Yuen-Zhou, Jacob J. Krich, and Alán Aspuru-Guzik}

\email{aspuru@chemistry.harvard.edu}

\selectlanguage{english}%

\address{Department of Chemistry and Chemical Biology, Harvard University,
Cambridge, MA 02138 }

\begin{abstract}
We report a conceptually straightforward witness that isolates coherent
electronic oscillations from their vibronic counterparts in nonlinear
optical spectra of molecular aggregates: Coherent oscillations as
a function of waiting time in broadband pump/broadband probe spectra
correspond to coherent electronic oscillations. Oscillations in individual
peaks of 2D electronic spectra do not necessarily yield this conclusion.
Our witness is simpler to implement than quantum process tomography
and potentially resolves a long-standing controversy on the character
of oscillations in ultrafast spectra of photosynthetic light harvesting
systems.
\end{abstract}

\maketitle

Recently, there has been considerable interest in long-lived quantum
superpositions of electronic states in photosynthetic molecular aggregates
and their potential role in efficient energy transport in biological
conditions \cite{mohseni,plenio}. Evidence for such electronic coherences
stems from time oscillations in peaks of two-dimensional electronic
spectra (2D-ES) which persist for over 600 fs \cite{engelfleming,scholes,engelchicago}.
However, coherences between vibronic levels involving a single electronic
state exhibit similar signatures in 2D-ES \cite{egorova,kaufmann_group,2012arXiv1201.6325C}
and have been shown to nontrivially affect energy transfer \cite{biggs_cina_anharmonic,chin_plenio,moran_vibrations}.
Although there are additional hints that support the interpretation
of the oscillations as due to electronic states (beating frequencies
and comparison with all-atom simulations \cite{Shim2012649}), unambiguous
tools to experimentally unravel the nature of these oscillations are
required. A big step has been the observation that, under weak coupling
to vibrations and negligible coherence transfer processes, electronic
coherences imply oscillations in off-diagonal peaks of rephasing 2D-ES
and in diagonal peaks of their non-rephasing counterparts \cite{cheng},
whereas general vibronic coherences show up as oscillations in any
region of either spectra \cite{turner}. However, the rephasing 2D-ES
of the paradigmatic Fenna-Matthews-Olson (FMO) complex exhibits oscillations
in both diagonal and off-diagonal peaks, indicating that systems of
interest may lie in the regime of strong coupling to vibrations \cite{Panitchayangkoon27122011}
or exhibit vibronic coherences only \cite{2012arXiv1201.6325C}. Techniques
of wavepacket reconstruction \cite{cinabiggs1} or quantum process
tomography (QPT) \cite{yuen-aspuru,yuenzhou} should clearly provide
an answer at a cost of several experiments. Our purpose here is to
provide a practical witness for coherent electronic oscillations,
which is applicable across different regimes of weak and strong coupling
to vibrations.

We illustrate the witness by considering the simplest molecular exciton
model, the coupled dimer \cite{forster}. Its Hamiltonian is given
by $H_{0}(\boldsymbol{R})=T_{N}+H_{el}(\boldsymbol{R})$, where $T_{N}$
is the nuclear kinetic energy, and $H_{el}(\boldsymbol{R})$ is the
electronic Hamiltonian which depends on the nuclei $\boldsymbol{R}$,
$H_{el}(\boldsymbol{R})=\sum_{mn}V_{mn}(\boldsymbol{R})|mn\rangle\langle mn|+J(\boldsymbol{R})(|10\rangle\langle01|+|01\rangle\langle10|).$
$|mn\rangle$ denotes the electronic state with $m$, $n$ excitations
in the first, second molecules, respectively ($m,n\in\{0,1\}$), $V_{mn}(\boldsymbol{R})$
is the corresponding diabatic potential energy surface, and $J(\boldsymbol{R})$
is the coupling between site excitations. Any pure state $|\Psi\rangle$
may be expressed in terms of vibronic states, that is, product states
of the electronic (system) and nuclear (bath) degrees of freedom,
$|\Psi\rangle=\sum_{i}a_{i}|e_{i}\rangle|N_{i}\rangle$, for coefficients
$a_{i}$, and $\{|e_{i}\rangle,|N_{i}\rangle\}$ electronic and nuclear
bases. A reduced electronic description of $|\Psi\rangle$ is obtained
by performing a trace over the bath, $\rho_{el}=\mbox{Tr}_{nuc}(|\Psi\rangle\langle\Psi|)$.
We consider light-matter perturbation in the dipole approximation,
$H_{pert}(s)=-\boldsymbol{\mu}\cdot\boldsymbol{\epsilon}(\boldsymbol{r},s)$,
where $\boldsymbol{\mu}=\sum_{e=01,10}\left(\boldsymbol{\mu}_{eg}|e\rangle\langle g|+\boldsymbol{\mu}_{fe}|f\rangle\langle e|\right)+\mbox{h.c.}$
is the dipole operator, and $\boldsymbol{\epsilon}(\boldsymbol{r},s)=\sum_{p=P,P'}[\epsilon_{p}(s-t_{p})\boldsymbol{e_{p}}+\mbox{c.c.}]$
denotes the pump (P) and probe (P') pulses, with $\epsilon_{p}(s-t_{p})=\frac{\lambda}{\sqrt{2\pi\sigma^{2}}}e^{-i\omega_{p}(s-t_{p})-(s-t_{p})^{2}/2\sigma^{2}}$
the Gaussian time-profile. Here, $\lambda$, $\omega_{p}$, $t_{p}$,
$\sigma$, and $\boldsymbol{e_{p}}$, are the strength, carrier frequency,
center time, width, and polarization of the $p$-th pulse, respectively.
We shall discuss PP' spectra $S_{PP'}(T)$ as a function of $T=t_{P'}-t_{P}$
(waiting time) \cite{mukamel}, which can be recovered from a 2D-ES
by integration along both frequency axes (Supplementary Material \cite{Supplemental material}
sec. I, SI-I). The main result of this article is: \emph{In the Condon
approximation and the broadband limit ($\sigma\to0$), oscillations
of }$S_{PP'}(T)$\emph{ as a function of $T$ correspond to coherent
electronic oscillations}; in this limit, $S_{PP'}(T)$ may be expressed
solely in terms of reduced electronic states \emph{$\rho_{el}$},
so oscillations cannot be due exclusively to nuclear dynamics.

The PP' signal may be written as the sum of $S_{SE}(T)$, $S_{ESA}(T)$,
and $S_{GSB}(T)$, with separate contributions from stimulated emission
(SE), excited state absorption (ESA), and ground state bleach (GSB)
\cite{minhaengbook}. If the initial vibrational state is known, each
of these terms may be expressed as a suitable wavefunction overlap
(SI-I \cite{Supplemental material}). For example, let the initial
wavefunction (before any pulse) be $|\Psi_{0}(0)\rangle=|g\rangle|\nu_{i}^{(g)}\rangle$,
where $|\nu_{i}^{(\eta)}\rangle$ is a vibrational eigenstate of $H_{vib,\eta}(\boldsymbol{R})\equiv T_{N}+V_{\eta}(\boldsymbol{R})$.
Treating the laser pulses pertubatively, the first order wavefunction
due to P is ($\hbar=1$) $|\Psi_{P}(s)\rangle=i\int_{-\infty}^{\infty}ds'e^{-iH_{0}(s-s')}\{\boldsymbol{\mu}\cdot\boldsymbol{\epsilon}_{P}(s'-t_{P})\}|\Psi_{0}(s')\rangle$,
and the second order wavefunction due to both P and P' is $|\Psi_{PP'}(s)\rangle=i\int_{-\infty}^{\infty}ds'e^{-iH_{0}(s-s')}\{\boldsymbol{\mu}\cdot\boldsymbol{\epsilon}_{P'}(s')\}|\Psi_{P}(s')\rangle$.
It can be shown that $S_{SE}(T)=\langle\Psi_{PP'}(s)|g\rangle\langle g|\Psi_{PP'}(s)\rangle$
(SI-I \cite{Supplemental material} and \cite{tannor,cina_fleming}).

\emph{Preliminary example.--- } We will develop some intuition through
an illustration, in which we focus on $S_{SE}(T)$. Consider the case
where the surfaces of the singly-excited diabatic states have the
same shape, $V_{10}(\boldsymbol{R})=V_{01}(\boldsymbol{R})+c$, for
some constant $c$ (but in general $V_{g}(\boldsymbol{R}),V_{f}(\boldsymbol{R})\neq V_{e}(\boldsymbol{R})+c$
for $e=01,10$). It is convenient to introduce the excitonic basis
$\{|g\rangle,|\alpha\rangle,|\beta\rangle,|f\rangle\}$, which diagonalizes
the electronic Hamiltonian at the ground state nuclear configuration:
$H_{el}(\boldsymbol{0})=\omega_{g}|g\rangle\langle g|+\omega_{\alpha}|\alpha\rangle\langle\alpha|+\omega_{\beta}|\beta\rangle\langle\beta|+\omega_{f}|f\rangle\langle f|$.
Here, $|g\rangle=|00\rangle$ and $|f\rangle=|11\rangle$, but in
general, $|\alpha\rangle$ and $|\beta\rangle$ differ from $|01\rangle$
and $|10\rangle$ in that they are delocalized due to $J(\boldsymbol{0})$.
Note that both $|\alpha\rangle$ and $|\beta\rangle$ are coupled
in the same way to the vibrational bath, and hence they form a decoherence-free
subspace \cite{lidar}. The first order wavefunction {}``right before''
the probe pulse may be expanded as $|\Psi_{P}(t_{1}+T)\rangle=\sum_{i=\alpha,\beta}\sum_{i}c_{i,m}(T)|i\rangle|\nu_{m}^{(i)}\rangle$.
Since in this case, $|i\rangle|\nu_{m}^{(i)}\rangle$ are eigenstates
of the molecular Hamiltonian $H_{0}(\boldsymbol{R})$, the excitons
are the adiabatic electronic states, there is no dissipation in the
electronic system, and the values $|c_{i,m}(T)|^{2}$ are constants
as a function of $T$, depending only on the details of $\mbox{P}$
\footnote{This is within the assumptions of the model; e.g., $T$ must be shorter
than the fluorescence timescale, $T<1\,\mbox{ns}$, approximately.%
}. The wavefunction {}``right after'' the probe at time $T$ is,
in the Condon approximation, given by, $|\Psi_{PP'}(t_{1}+T)\rangle=i\sum_{i=\alpha,\beta}\sum_{mn}\boldsymbol{\mu}_{mg}\cdot\mbox{\ensuremath{\boldsymbol{e}}}_{P'}\tilde{\epsilon}_{P'}(\omega_{im,gn})\langle\nu_{n}^{(g)}|\nu_{m}^{(i)}\rangle c_{i,m}(T)|g\rangle|\nu_{n}^{(g)}\rangle$,
where $\tilde{\epsilon}_{p}(\omega)=\lambda e^{-(\omega-\omega_{p})^{2}\sigma^{2}/2}$
is the Fourier transform of pulse $p$ at frequency $\omega$. This
expression can be interpreted as a wavepacket in the ground state
created when the probe couples the vibrational levels of the singly-excited
states to the vibrational levels of the ground state via the electric
dipole moment, where the amplitudes in the various vibrational levels
depends on the probe's electric field at the given transition energy
and the Condon overlap. Computing the norm of the resulting wavepacket,

\begin{eqnarray}
S_{SE}(T) & = & \sum_{ij=\alpha,\beta}(\boldsymbol{\mu}_{ig}\cdot\boldsymbol{e}_{P'})(\boldsymbol{\mu}_{jg}\cdot\boldsymbol{e}_{P'})\nonumber \\
 &  & \times\sum_{mm'n}\langle\nu_{m'}^{(j)}|\nu_{n}^{(g)}\rangle\langle\nu_{n}^{(g)}|\nu_{m}^{(i)}\rangle\nonumber \\
 &  & \times\tilde{\epsilon}_{P'}(\omega_{im,gn})\tilde{\epsilon}_{P'}^{*}(\omega_{jm',gn})c_{i,m}(T)c_{j,m'}^{*}(T)\},\label{eq:PP narrowband}
\end{eqnarray}
which corresponds to sums of interferences between vibrational states
of the same and different excitonic states, respectively, projecting
onto the same vibrational state in the ground state. Note that $S_{SE}(T)$
can be written as a linear combination of elements of the full vibronic
density matrix $\rho(T)=|\Psi_{P}(t_{1}+T)\rangle\langle\Psi_{P}(t_{1}+T)|$.
The terms $\langle i,m|\rho(T)|j,m'\rangle=c_{im}(T)c_{jm'}^{*}(T)$
for $(i,m)\neq(j',m')$ correspond to \emph{vibronic coherences} and
oscillate at the difference frequency between the $|i\rangle|m\rangle$
and the $|j\rangle|m'\rangle$ states. When we consider the broadband
(\emph{bb}) limit of Eq.\ (\ref{eq:PP narrowband}) , where $\tilde{\epsilon}(\omega)=\lambda$
for all the $\omega$ values of interest,

\begin{eqnarray}
S_{SE}^{bb}(T) & = & \lambda^{2}\sum_{ij=\alpha,\beta}(\boldsymbol{\mu}_{ig}\cdot\boldsymbol{e}{}_{P'})(\boldsymbol{\mu}_{jg}\cdot\boldsymbol{e}_{P'})\sum_{m}c_{im}(T)c_{jm}^{*}(T).\label{eq:PP broadband}
\end{eqnarray}

Crucially, Eq.\ (\ref{eq:PP broadband}) is a linear combination
of elements of $\rho_{el}(T)$ as opposed to the full vibronic space.
In fact, the terms for $i=j$ correspond to electronic populations
and, due to the absence of electronic decoherence in this example,
stay constant with respect to $T$. The term $\langle\alpha|\rho_{el}(T)|\beta\rangle=\sum_{i}c_{i,\alpha}(T)c_{i,\beta}^{*}(T)$
corresponds to an electronic coherence between $|\alpha\rangle$ and
$|\beta\rangle$, and shows oscillations at the single frequency $\omega_{\alpha\beta}$
as a function of $T$. Hence, coherent oscillations in $S_{SE}^{bb}(T)$
are a witness for coherent electronic dynamics. Remarkably, in the
additional limit where one of the excitons is dark (e.g., $\boldsymbol{\mu}_{\beta g}=0$),
we have a monomer instead of a dimer, and $S_{SE}^{bb}(T)$ is a constant
even in the case of large Condon displacements, where there is large
vibrational motion between pump and probe. This observation for the
monomer has been previously reported by Yan and Mukamel \cite{yan_mukamel}.

The results above can be interpreted as follows. In the Condon approximation,
the probe couples only to the electronic dipole, so in the broadband
limit it acts uniformly across every transition energy, and hence
across every nuclear configuration within a particular electronic
state. In general, $S_{SE}(T)$ is a sum of multiple interferences
among portions of wavepackets at different electronic and nuclear
configurations. In $S_{SE}^{bb}(T)$, the probe opens only two interference
pathways (just as in the double-slit experiment), via emission from
the $|\alpha\rangle$ or the $|\beta\rangle$ state, insensitive to
vibrational dynamics, providing a witness for coherent electronic
oscillations.

\emph{General case.---} The example above readily generalizes to include
effects of initial thermalized states of the bath, ESA and GSB contributions,
and non-adiabatic effects. In the limit of broadband $\mbox{P}$ (SI-II
and III, \cite{Supplemental material}) and $\mbox{P}'$, each of
the contributions to $S_{PP'}^{bb}(T)$ are (SI-II, \cite{Supplemental material}),
$S_{SE}^{bb}(T)=\lambda^{4}\sum_{ijpq}(\boldsymbol{\mu}_{gi}\cdot\boldsymbol{e}_{P'})(\boldsymbol{\mu}_{qg}\cdot\mbox{\ensuremath{\boldsymbol{e}}}_{P})(\boldsymbol{\mu}_{gp}\cdot\mbox{\ensuremath{\boldsymbol{e}}}_{P})(\boldsymbol{\mu}_{jg}\cdot\boldsymbol{e}_{P'})\chi_{ijqp}(T)$,
$S_{ESA}^{bb}(T)=-\lambda^{4}\sum_{ijpq}(\boldsymbol{\mu}_{fi}\cdot\boldsymbol{e}_{P'})(\boldsymbol{\mu}_{qg}\cdot\mbox{\ensuremath{\boldsymbol{e}}}_{P})(\boldsymbol{\mu}_{gp}\cdot\mbox{\ensuremath{\boldsymbol{e}}}_{P})(\boldsymbol{\mu}_{jf}\cdot\boldsymbol{e}_{P'})\chi_{ijqp}(T)$,
and $S_{GSB}^{bb}(T)=\lambda^{4}\sum_{ip}(\boldsymbol{\mu}_{gp}\cdot\mbox{\ensuremath{\boldsymbol{e}}}_{P})(\boldsymbol{\mu}_{pg}\cdot\mbox{\ensuremath{\boldsymbol{e}}}_{P})(\boldsymbol{\mu}_{gi}\cdot\boldsymbol{e}_{P'})(\boldsymbol{\mu}_{ig}\cdot\boldsymbol{e}_{P'})$,
where the \emph{process matrix} $\chi(T)$ is given by, $\chi_{ijqp}(T)=\mbox{Tr}_{nuc}\{\langle i|e^{-iH_{0}T}\left(|q\rangle\langle p|\otimes\rho_{B}(0)\right)e^{iH_{0}T}|j\rangle\}$
\cite{yuen-aspuru,yuenzhou}, and it is easy to see that $S_{PP'}^{bb}(T)$
is invariant under change of electronic basis within the singly-exicted
states. Here, $\rho_{B}(0)=\sum_{n}p_{n}|\nu_{n}^{(g)}\rangle\langle\nu_{n}^{(g)}|$
is the initial thermal vibrational ensemble in the ground electronic
state. $\chi(T)$ describes the evolution of the electronic system,
assuming that the vibrational system begins in $\rho_{B}(0)$. If
the initial state of the bath can be prepared at $\rho_{B}(0)$ regardless
of the electronic state, as in the impulsive limit, an integrated
equation of motion can be written as $\rho_{ij}(T)=\sum_{ijqp}\chi_{ijqp}(T)\rho_{qp}(0)$.
As in the preliminary example, $S_{PP'}^{bb}(T)$ is a linear combination
of entries of reduced states $\rho_{el}(T)$, so oscillations in it
are a manifestation of electronic oscillations, justifying the witness.

Given an electronic basis, any element $\chi_{ijqp}(T)$ can in principle
exhibit oscillations. For a large variety of systems, it is however,
possible to associate the largest amplitude oscillations of $\chi(T)$
to electronic coherences in some basis. In the preliminary example,
the lack of dissipation implies that $\chi_{ijqp}(T)=\delta_{iq}\delta_{jp}e^{-i\omega_{qp}T}$,
so the only possible oscillatory contribution to $S_{PP'}(T)$ corresponds
to $\chi_{\alpha\beta\alpha\beta}(T)=\chi_{\beta\alpha\beta\alpha}^{*}(T)$
(excitonic coherence). In the non-adiabatic case where $V_{01}(\boldsymbol{R})\neq V_{10}(\boldsymbol{R})+c$,
each electronic state couples differently to the vibrational modes.
However, in the limit of weak system-bath coupling, the vibronic states
$|e\rangle|\nu_{j}^{(e)}\rangle$ are still the correct eigenstates
of $H_{0}(\boldsymbol{R})$ up to zeroth order in the coupling, so
any oscillations in the signal will still be dominated by excitonic
coherences. Finally, for intermediate and strong system-bath coupling
together with a fast bath decorrelation timescale, a polaron transformation
defines an electronic basis $\{|g\rangle,|\tilde{\alpha}\rangle,|\tilde{\beta}\rangle,|f\rangle\}$
that diagonalizes a zeroth-order electronic Hamiltonian weakly coupled
to a renormalized bath (\cite{silbey-harris} and SI-IV \cite{Supplemental material}).
In this case, the highest amplitude oscillations in its $S_{PP'}^{bb}(T)$
would correspond to electronic coherences $\chi_{\tilde{\alpha}\tilde{\beta}\tilde{\alpha}\tilde{\beta}}(T)=\chi_{\tilde{\beta}\tilde{\alpha}\tilde{\beta}\tilde{\alpha}}^{*}(T)$.
For more general aggregates, if this were an issue of interest, a
partial QPT could be designed to determine the value of specific terms
of $\chi(T)$ \cite{yuen-aspuru,yuenzhou}.

\emph{Numerical examples.--- }We have performed simulations for a
monomer, a dimer which exhibits coherent electronic oscillations,
and an incoherent dimer, where each singly-excited site is coupled
to a single vibrational mode. These three examples illustrate the
value of the witness (Fig. 1), as all three have oscillatory 2D-ES
(Fig. 2), but the monomer and incoherent dimer do not have coherent
electronic oscillations. The witness correctly shows that only the
coherent dimer has a positive witness. The simulations include inhomogeneous
broadening (ensembles of 500 molecules with Gaussian site disorder
of standard deviation $40$ cm$^{-1}$ and, for the dimers, site energy
correlation 0.8), thermal averaging of initial vibrational states
according to a Boltzmann distribution at 273 K, isotropic averaging,
and explicit inclusion of pulses with the dynamics. Roughly, there
are two energy scales to consider, an average coupling $J$ and a
reorganization energy $\lambda$, in which case the impulsive limit
is set by $\frac{1}{\sigma}\gg\mbox{max}(J,\lambda)$. For these simulations,
the pulses are within the FWHM=10--20 fs range, and cover the entire
absorption spectra, respectively (SI-V, \cite{Supplemental material}).
Fig.\ 1 shows $\langle S_{PP'}^{bb}(T)\rangle_{zzzz}$, the witness
averaged at the collinear pulse setting $zzzz$, for about 900 fs
(top). We can associate the witness oscillations to oscillations of
elements in $\chi(T)$. We show a few representative elements of this
matrix (bottom). Fig.\ 2 presents snapshots of the rephasing 2D-ES,
$\langle\tilde{S}(\omega_{\tau},T,\omega_{t})\rangle_{zzzz}$, for
a sampling of waiting times $T$ between 71.6 and 270.6 fs (left),
indicating that vibronic coherences manifest as diagonal and cross-peak
oscillations %
\footnote{Inhomogeneous broadening was not included in the 2D-ES due to the
expensive cost of their computation.%
}. Notice that due to strong coupling to vibrations, the coherent dimer
also exhibits oscillations in the diagonal peaks, implying the inapplicability
of previous measures for this case \cite{cheng,turner}. As another
illustration, the integrated signal under the cross-peaks encircled
in black is in the right plots. Note that the largest amplitude oscillations
are in the monomer, which cannot have coherent electronic oscillations,
showing that oscillations in peaks in the 2D-ES do not directly translate
into coherent electronic dynamics, and hence are not the correct witness.

The witness is positive if, once the dc background is subtracted from
$S_{PP'}^{bb}(T)$, there are oscillations with amplitude proportional
to $\mu^{4}$, where $\mu$ is some estimate of an \emph{electronic}
transition dipole moment. If spurious oscillations due to finite pulse-duration
are suspected, a more quantitative confirmation is the following:
(a) Collect traces $S_{PP}^{bb}(T)$ at several pulse widths $\sigma$,
all roughly in the broadband domain. (b) Fourier transform the data:
$\tilde{S}_{PP}^{bb}(\omega_{T})=\frac{1}{2\pi}\int_{0}^{\infty}dTe^{i\omega T}S_{PP}^{bb}(T)$.
(c) Locate non-zero frequencies of $\tilde{S}_{PP}^{bb}(\omega_{T})$
corresponding to oscillations between discrete states (ignore the
dc component). For each of these frequencies $\omega_{T}$, plot $\tilde{S}_{PP}^{bb}(\omega_{T})$
as a function of $\sigma$, and linearly extrapolate to $\sigma\to0$.
If the obtained intercepts are at zero within noise levels, the witness
is negative . SI-II \cite{Supplemental material} displays an analytical
expression for the $O(\sigma)$ correction of $S_{PP}^{bb}(T)$, providing
a theoretical basis for this procedure.

Although the theory has been detailed here for a dimer, the witness
is applicable to larger aggregates. In the case of FMO, due to spectral
congestion, it might be fruitful to focus on pairs of exciton states
at a time, for instance, the first and the third exciton states, either
via direct PP' measurements that cover these transitions exclusively,
or alternatively, integrating windows of broadband 2D-ES corresponding
to these two states only, assuming that relaxation processes do not
occur outside of this spectral window.

We dedicate this letter to the late Bob Silbey who at different stages
introduced J.Y.Z. to theoretical chemistry and mentored A.A.G in the
field. We thank DARPA Award No. N66001-10-4060 and EFRC-DOE Award
No. DE-SC0001088.

\begin{figure}
\begin{centering}
\includegraphics[width=8.5cm]{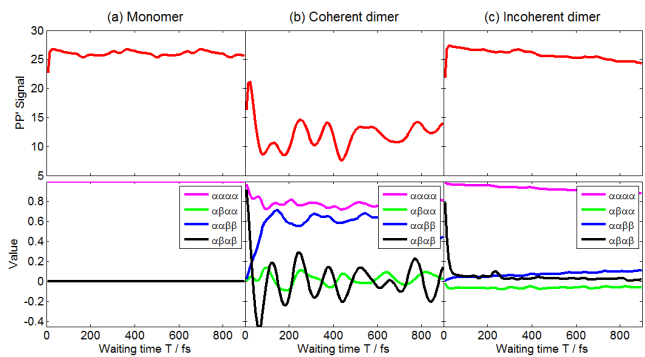}
\par\end{centering}

\caption{(Top) Broadband PP' spectra as a function of waiting time $T$ as
a witness for coherent electronic oscillations. The small oscillations
in (a) and (c) are due to finite pulse durations. (Bottom) The witness
is a linear combination of elements of the process matrix $\chi(T)$.
Traces of a few representative elements of $\chi(T)$ are displayed.}
\end{figure}

\begin{figure}
\begin{centering}
\includegraphics[width=8cm]{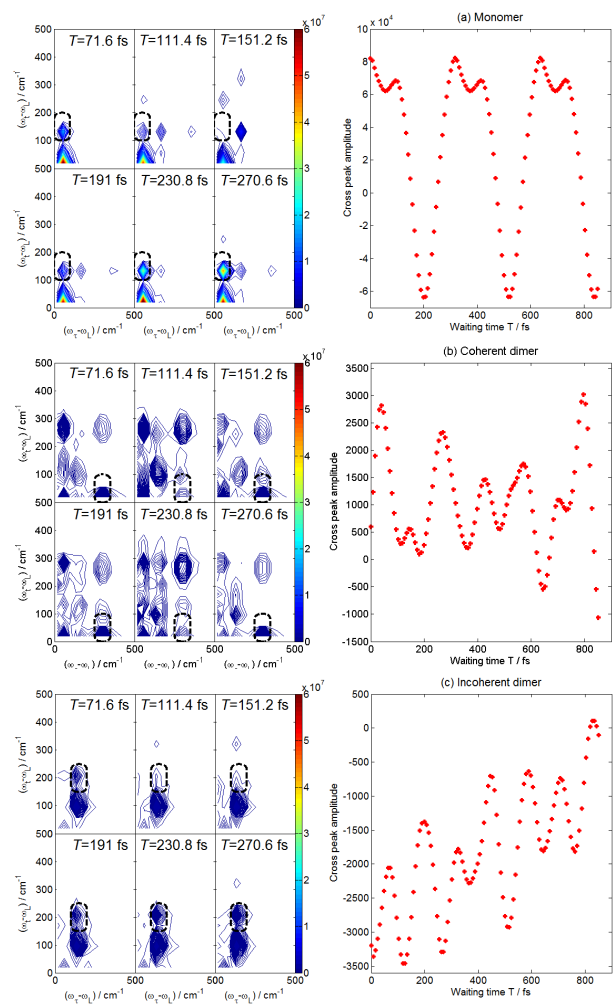}
\par\end{centering}

\caption{(Left) Norm of broadband rephasing 2D-ES for (a) monomer, (b) coherent
dimer, and (c) incoherent dimer. Diagonal and cross peaks oscillate
as a function of time in all cases, indicating general vibronic coherences
but not necessarily electronic coherence. Color scale (arbitrary units)
is fixed throughout. Black dotted circles at left indicate cross peaks
whose real part amplitude is shown followed in a finer time grid in
the right panels (varying axes for each sample, arbitrary units).
These traces are the wrong witness for coherent electronic oscillations.}
\end{figure}

\bibliographystyle{unsrt}

\pagebreak{}

\onecolumngrid

\section*{\setcounter{page}{1}}

\global\long\def\theequation{S\arabic{equation}}
 \setcounter{equation}{0}

\global\long\def\thefigure{S\arabic{figure}}
 \setcounter{figure}{0}

\begin{center}
\Large{\emph{Supplementary Information} }
\end{center}
%
%
%
%
%
%

\section{PP' signal in terms of wavepacket overlaps}

Consider the situation described in the article, where the total Hamiltonian
is given by $H=H_{0}(\boldsymbol{R})+H_{pert}(s)$. In this section,
we will assume $H_{0}(\boldsymbol{R})$ to be the same molecular piece
as the one described in the article, and $H_{pert}(s)=-\boldsymbol{\mu}\cdot\boldsymbol{\epsilon}(\boldsymbol{r},s)$
to be the standard light-matter interaction in the dipole approximation,
although we consider a slightly more general setup, where the electric
field is described by three non-collinear beams, $\boldsymbol{\epsilon}(\boldsymbol{r},s)=\sum_{p=1}^{3}[\epsilon_{p}(s-t_{p})e^{i\boldsymbol{k}_{p}\cdot\boldsymbol{r}+i\phi_{p}}\boldsymbol{e_{p}}+\mbox{c.c.}]$
with different wavevectors $\boldsymbol{k}_{p}$ and phases $\phi_{p}$.
The expressions for $S_{PP'}(T)$ will appear as we take the limit
of the PE signal to the PP' limit.

The pulses generate a time-dependent polarization $\boldsymbol{P}(\boldsymbol{r},s)=\mbox{Tr}(\boldsymbol{\mu}\rho(\boldsymbol{r}))=\sum_{\boldsymbol{k}}P(\boldsymbol{k};s)e^{i\boldsymbol{k}_{s}\cdot\boldsymbol{r}}$
on each molecule at position $\boldsymbol{r}$ %
\footnote{We use the word \emph{polarization} in two different ways: To denote
(a) the orientation of oscillations of the electric field and (b)
the density of electric dipole moments in a material. The meaning
should be clear by the context. %
}. The allowed wavevectors are the phase-matching directions $\boldsymbol{k}=q\boldsymbol{k}_{1}+r\boldsymbol{k}_{2}+s\boldsymbol{k}_{3}$
for integers $q$, $r$, $s$, and encode different sequences of interactions
of the pulses with the molecule. We are interested in the signal $S$
at the photon-echo (PE) phase-matched direction $\boldsymbol{k}_{PE}=-\boldsymbol{k}_{1}+\boldsymbol{k}_{2}+\boldsymbol{k}_{3}$,
which can be detected by mixing the material ensemble emission with
a local oscillator (LO) pulse $\boldsymbol{\epsilon}_{4}(s)$ travelling
along $\boldsymbol{k}_{4}=\boldsymbol{k}_{PE}$, $S(\tau,T,t)=-2\Im\int_{-\infty}^{\infty}ds\epsilon_{4}^{*}(s-t_{4})\mbox{\ensuremath{\boldsymbol{e}}}_{4}\cdot\boldsymbol{P}(\boldsymbol{k}_{PE};\tau,T,s)$,
where $\tau=t_{2}-t_{1}$ (coherence time), $T=t_{3}-t_{2}$ (waiting
time), and $t=t_{4}-t_{3}$ (echo time) \cite{mukamel}. Upon repeated
collection of $S(\tau,T,t)$ for many values of time intervals, a
2D-ES can be constructed as a function of $T$, by Fourier transforming
the signal with respect to $\tau$ and $t$, $\tilde{S}(\omega_{\tau},T,\omega_{t})=\int_{0}^{\infty}d\tau e^{-i\omega_{\tau}\tau}\int_{0}^{\infty}dte^{i\omega_{t}t}S(\tau,T,t)$
\cite{mukamel_2d,minhaengbook}. In general, oscillations in $S(\tau,T,t)$
and $\tilde{S}(\omega_{\tau},T,\omega_{t})$ can be associated to
coherent superpositions of vibronic eigenstates of $H_{0}(\boldsymbol{R})$,
but not necessarily of electronic states \cite{egorova}. In the article,
we paid special attention to the pump-probe (PP) limit $S_{PP'}(T)$,
which is equivalent to a \emph{differential transmission} signal,
where the first two pulses act as the pump P, ($\epsilon_{1}=\epsilon_{2}\equiv\epsilon_{P}$,
$\phi_{1}=\phi_{2}\equiv\phi_{P}$), the last two as the probe P'
($\epsilon_{3}=\epsilon_{4}=\epsilon_{P'}$ and $\phi_{3}=\phi_{4}\equiv\phi_{P'}$),
$\tau=t=0$, and P and P' are well separated (i.e., $T\gg\sigma$).
$S_{PP'}(T)$ can be recovered from the 2D-ES as an inverse Fourier
transform at zero frequencies, $S_{PP'}(T)\equiv S_{PE}(0,T,0)=\frac{1}{(2\pi)^{2}}\int_{-\infty}^{\infty}d\omega_{\tau}\int_{-\infty}^{\infty}d\omega_{t}\tilde{S}(\omega_{\tau},T,\omega_{t})$.
This limit justifies the form of $H_{pert}(s)$ given in the article,
which only consists of two pulses.

The starting point is the expression for $S_{PP'}(T)$,

\begin{equation}
S_{PP'}(T)=-2\Im\int_{-\infty}^{\infty}ds'\epsilon_{4}^{*}(s'-t_{4})\mbox{\ensuremath{\boldsymbol{e}}}_{4}\cdot\boldsymbol{P}(\boldsymbol{k}_{PE};0,T,0).\label{eq:S_PP'}
\end{equation}

We shall derive a wavepacket overlap formula for $S_{PP'}(T)$ assuming
that P and P' are well separated, $T\gg\sigma$, analogously to the
doorway-window approach \cite{mukamel}. First, we conveniently define
the following wavefunctions:
\begin{eqnarray}
|\Psi_{0}(s)\rangle & = & e^{-iH_{0}(s-t_{P})}|\Psi_{0}(s')\rangle,\label{eq:0}\\
|\Psi_{P}(s)\rangle & = & i\int_{-\infty}^{\infty}ds'e^{-iH_{0}(s-s')}\{\boldsymbol{\mu}\cdot\mbox{\ensuremath{\boldsymbol{e}}}_{P}(\epsilon_{P}(s'-t_{P})+\mbox{c.c.})\}|\Psi_{0}(s')\rangle,\label{eq:P}\\
|\Psi_{PP'}(s)\rangle & = & i\int_{-\infty}^{\infty}ds'e^{-iH_{0}(s-s')}\{\boldsymbol{\mu}\cdot\mbox{\ensuremath{\boldsymbol{e}}}_{P'}(\epsilon_{P'}(s'-t_{P'})+\mbox{c.c.})\}|\Psi_{P'}(s')\rangle,\label{eq:PP'}\\
|\Psi_{PP}(s)\rangle & = & (i)^{2}\int_{-\infty}^{\infty}ds'\int_{-\infty}^{s'}ds''e^{-iH_{0}(s-s')}\{\boldsymbol{\mu}\cdot\mbox{\ensuremath{\boldsymbol{e}}}_{P}(\epsilon_{P}(s'-t_{P})+\mbox{c.c.})\}\label{eq:PP}\\
 &  & \times e^{-iH_{0}(s'-s'')}\{\boldsymbol{\mu}\cdot\mbox{\ensuremath{\boldsymbol{e}}}_{P}(\epsilon_{P}(s''-t_{P})+\mbox{c.c.})\}|\Psi_{0}(s')\rangle,\nonumber \\
|\Psi_{P'P'}(s)\rangle & = & (i)^{2}\int_{-\infty}^{\infty}ds'\int_{-\infty}^{s'}ds''e^{-iH_{0}(s-s')}\{\boldsymbol{\mu}\cdot\mbox{\ensuremath{\boldsymbol{e}}}_{P'}(\epsilon_{P'}(s'-t_{P'})+\mbox{c.c.})\}\label{eq:P'P'}\\
 &  & \times e^{-iH_{0}(s'-s'')}\{\boldsymbol{\mu}\cdot\mbox{\ensuremath{\boldsymbol{e}}}_{P'}(\epsilon_{P'}(s''-t_{P'})+\mbox{c.c.})\}|\Psi_{0}(s'')\rangle,\nonumber \\
|\Psi_{PPP'P'}(s)\rangle & = & (i)^{2}\int_{-\infty}^{\infty}ds'\int_{-\infty}^{s'}ds''e^{-iH_{0}(s-s')}\{\boldsymbol{\mu}\cdot\mbox{\ensuremath{\boldsymbol{e}}}_{P'}(\epsilon_{P'}(s'-t_{P'})+\mbox{c.c.})\}\label{eq:PPP'P'}\\
 &  & \times e^{-iH_{0}(s'-s'')}\{\boldsymbol{\mu}\cdot\mbox{\ensuremath{\boldsymbol{e}}}_{P'}(\epsilon_{P'}(s''-t_{P'})+\mbox{c.c.})\}|\Psi_{PP}(s'')\rangle,\nonumber
\end{eqnarray}
 which are valid for $s\gg t_{P'}$ (after the envelopes of the pulses
have considerably decayed), and which correspond to the processes
indicated by their subscripts, i.e., $|\Psi_{PPP'P'}(s)\rangle$ corresponds
to the fourth order wavefunction ($O(\lambda^{4})$) resulting from
two actions of P and two of P' (see Fig. S1).

Eqs. (\ref{eq:0})--(\ref{eq:PPP'P'}) allow for a calculation of
$\boldsymbol{P}(\boldsymbol{k}_{PE};0,T,0)$ and hence of $S_{PP'}(T)$
via Eq. (\ref{eq:S_PP'}). Note that, as opposed to a general PE signal,
$S_{PP'}(T)$ does not depend on the phases of the pulses because
$\phi_{1}=\phi_{2}\equiv\phi_{P}$ and $\phi_{3}=\phi_{4}\equiv\phi_{P'}$.
The phase-matching condition $\boldsymbol{k}_{PE}=-\boldsymbol{k}_{1}+\boldsymbol{k}_{2}+\boldsymbol{k}_{3}$
together with the rotating-wave approximation indicate that for each
wavevector $+\boldsymbol{k}_{j}(-\boldsymbol{k}_{j})$, the pulse
$j$ acts with the term $\epsilon_{j}(\epsilon_{j}^{*})$, exciting
(de-exciting) the ket or de-exciting (exciting) the bra. Collecting
all the terms result in $S_{PP'}(T)=S_{SE}(T)+S_{ESA}(T)+S_{GSB}(T)$:

\begin{eqnarray}
S_{SE}(T) & = & -2\Im\int_{-\infty}^{\infty}dt'\langle\Psi_{PP'}(t')|\{\epsilon_{P'}^{*}(t'-t_{P'})\boldsymbol{\mu}\cdot\boldsymbol{e}_{P'}\}|\Psi_{P}(t')\rangle\nonumber \\
 & = & -2\Im(-i)\int_{-\infty}^{\infty}dt'\int_{-\infty}^{t'}ds'\langle\Psi_{P}(s')|\{-i\epsilon_{P'}(s'-t_{P'})\boldsymbol{\mu}\cdot\boldsymbol{e}_{P'}\}e^{iH_{0}(t'-s')}\{(i)\epsilon_{P'}^{*}(t'-t_{P'})\boldsymbol{\mu}\cdot\boldsymbol{e}_{P'}\}|\Psi_{P}(t')\rangle\nonumber \\
 & = & 2\langle\Psi_{PP'}(s)|g\rangle\langle g|\Psi_{PP'}(s)\rangle,\label{eq:SE_overlap}\\
S_{ESA}(T) & = & -2\Im\int_{-\infty}^{\infty}dt'\langle\Psi_{P}(t')|\{\epsilon_{P'}^{*}(t'-t_{P'})\boldsymbol{\mu}\cdot\boldsymbol{e}_{P'}\}|\Psi_{PP'}(t')\rangle\nonumber \\
 & = & -2\Im(i)\int_{-\infty}^{\infty}dt'\int_{-\infty}^{t'}ds'\langle\Psi_{P}(t')|\{(-i)\epsilon_{P'}^{*}(t'-t_{P'})\boldsymbol{\mu}\cdot\boldsymbol{e}_{P}\}e^{-iH_{0}(t'-s')}\{i\epsilon_{P}(s'-t_{P})\boldsymbol{\mu}\cdot\boldsymbol{e}_{P}\}|\Psi_{P}(s')\rangle\nonumber \\
 & = & -2\langle\Psi_{PP'}(s)|f\rangle\langle f|\Psi_{PP'}(s)\rangle,\label{eq:ESA_overlap}\\
S_{GSB}(T) & = & -2\Im\int_{-\infty}^{\infty}dt'\{\langle\Psi_{PP}(t')|\{\epsilon_{P'}^{*}(t'-t_{P'})\boldsymbol{\mu}\cdot\boldsymbol{e}_{P'}\}|\Psi_{P'}(t')\rangle+\langle\Psi_{0}(t')|\{\epsilon_{P'}^{*}(t'-t_{P'})\boldsymbol{\mu}\cdot\boldsymbol{e}_{P'}\}|\Psi_{PPP'}(t')\rangle\}\nonumber \\
 & = & -2\Im(-i)\int_{-\infty}^{\infty}dt'\{\langle\Psi_{PP}(t')|\{(i)\epsilon_{P'}^{*}(t'-t_{P'})\boldsymbol{\mu}\cdot\boldsymbol{e}_{P'}\}|\Psi_{P'}(t')\rangle+\langle\Psi_{0}(t')|\{(i)\epsilon_{P'}^{*}(t'-t_{P'})\boldsymbol{\mu}\cdot\boldsymbol{e}_{P'}\}|\Psi_{PPP'}(t')\rangle\}\nonumber \\
 & = & 2\Re\{\langle\Psi_{PP}(s)|g\rangle\langle g|\Psi_{P'P'}(s)\rangle+\langle\Psi_{0}(s)|g\rangle\langle g|\Psi_{PPP'P'}(s)\rangle\}.\label{eq:GSB_overlap}
\end{eqnarray}
 where again, $s\gg t_{P'}$, and otherwise, the signals are independent
of $s$. This can be easily understood in physical terms: After the
action of the pulses, the wavefunctions still evolve according to
$H_{0}(\boldsymbol{R})$, but the overlaps do not change in time.
Eqs. (\ref{eq:SE_overlap})---(\ref{eq:GSB_overlap}) are in the spirit
of wavepacket approaches to PP' spectroscopy \cite{mukamel_adv,cina_fleming,cinabiggs1,cinabiggs2,biggs_cina_anharmonic,cao_wilson,soo}.

In order to gain additional insight, we interpret the formulas in
terms of differential transmission by enumerating all the possible
absorption and emission processes which are quadratic in P and P'.
P promotes a wavepacket from $|g\rangle$ to $|\Psi_{P}(s)\rangle$,
a superposition of wavepackets in $|\alpha\rangle$ and $|\beta\rangle$.
P' acts on this state, creating $|\Psi_{PP'}(s)\rangle$, a superposition
of wavepackets in $|g\rangle$ and $|f\rangle$. Naturally, the photons
emitted in SE correspond to the squared amplitude of $\langle g|\Psi_{PP'}(t)\rangle$,
whereas the ones absorbed in ESA are associated with the squared amplitude
of $\langle f|\Psi_{PP'}(s)\rangle$, hence providing an intuition
for the expressions for $S_{SE}(T)$ and $S_{ESA}(T)$. Finally, $S_{GSB}(T)$
can be thought as accounting for the {}``leftover'' SE processes,
namely, overlaps between wavepackets created by pulses at different
times. After P and P' , the total ground state wavepacket is $\langle g|\Psi(t)\rangle=\langle g|(|\Psi_{0}(t)\rangle+|\Psi_{PP'}(t)\rangle+|\Psi_{PP}(t)\rangle+|\Psi_{P'P'}(t)\rangle+|\Psi_{PPP'P'}(t)\rangle+\mbox{higher order contributions})$.
Collecting wavepacket overlaps which are quadratic in both pulses
yields $S_{SE}(T)+S_{GSB}(T)$. \textquotedbl{}Leftover\textquotedbl{}
ESA processes do not contribute to $S_{PP'}(T)$ because they do not
fullfill the PE phase-matching condition (they appear in double-quantum
coherence spectroscopy, for instance).

Thermal effects follow from averaging the signals corresponding to
initial states $|\Psi_{0}(t_{P})\rangle$ sampled according to a Boltzmann
distribution.
\begin{figure}
\begin{centering}
\includegraphics[scale=0.45]{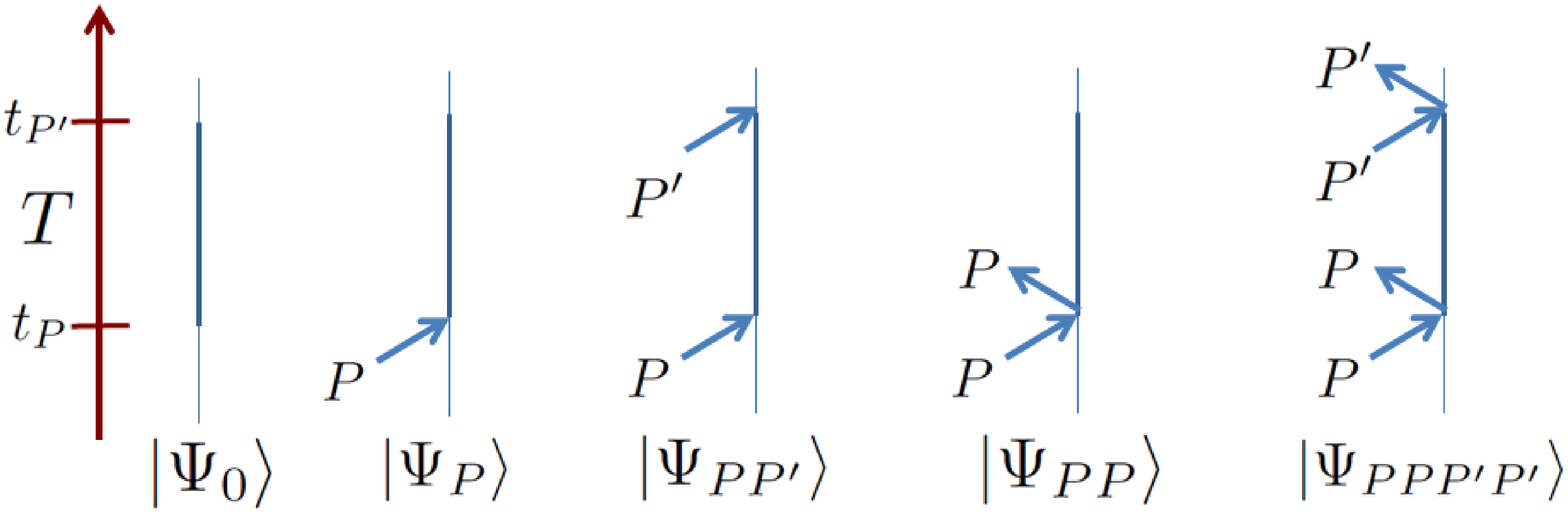}
\par\end{centering}

\caption{Feynman diagrams for the wavefunctions defined in Eqs. (\ref{eq:0})--(\ref{eq:PPP'P'}).}
\end{figure}

\section{General expressions for $S_{PP'}(T)$ in vibronic basis}

In order to manipulate the wavepacket overlap expressions, it is convenient
to define two vibronic bases:
\begin{itemize}
\item The vibronic eigenbasis of $H_{0}(\boldsymbol{R})$,$\{|g,\nu_{n}^{(g)}\rangle,|\zeta\rangle,|f,\nu_{n}^{(f)}\rangle\}$,
which satisfy

\begin{eqnarray}
H_{0}(\boldsymbol{R})|g,\nu_{n}^{(g)}\rangle & = & \omega_{gn}|\zeta\rangle,\label{eq:vibronic_eigenbasis_g}\\
H_{0}(\boldsymbol{R})|\zeta\rangle & = & \omega_{\zeta}|\zeta\rangle,\label{eq:vibronic_eigenbasis_zeta}\\
H_{0}(\boldsymbol{R})|f,\nu_{n}^{(f)}\rangle & = & \omega_{fn}|f,\nu_{n}^{(f)}\rangle.\label{eq:vibronic_eigenbasis_f}
\end{eqnarray}
 where $|\zeta\rangle$ corresponds to the singly excited state manifold.

\item The tensor product basis $\{|m,\nu_{n}^{(g)}\rangle\}$, where $\{|m\rangle\}$
denotes electronic states in an arbitrary electronic basis (for instance,
the excitonic one), and $\{|\nu_{n}^{(g)}\rangle\}$ refers to vibrational
eigenstates of the ground vibrational Hamiltonian, $H_{vib,00}(\boldsymbol{R})=T_{N}+V_{00}(\boldsymbol{R})$.
\end{itemize}
Note that we can always write states in the vibronic eigenbasis in
terms of the second one: $|g,\nu_{n}^{(g)}\rangle$ stays the same,
$|\zeta\rangle=\sum_{mn}\langle m,\nu_{n}^{(g)}|\zeta\rangle|m,\nu_{n}^{(g)}\rangle$,
and $|f,\nu_{n'}^{(f)}\rangle=\sum_{n}|f,\nu_{n'}^{(g)}\rangle\langle f,\nu_{n}^{(g)}|f,\nu_{n'}^{(f)}\rangle=\sum_{n}\langle\nu_{n}^{(g)}|\nu_{n'}^{(f)}\rangle|f,\nu_{n'}^{(g)}\rangle$.

Using both bases, the process matrix $\chi(T)$ affords a compact
representation,

\begin{eqnarray}
\chi_{ijqp}(T) & = & \chi_{ijqp}(T)=\mbox{Tr}_{nuc}\{\langle i|e^{-iH_{0}T}\left(|q\rangle\langle p|\otimes\rho_{B}(0)\right)e^{iH_{0}T}|j\rangle\}\nonumber \\
 & = & \sum_{\zeta\zeta'nn'}p_{n}e^{-i(\omega_{\zeta}-\omega_{\zeta'})T}\langle i,\nu_{n'}^{(g)}|\zeta\rangle\langle\zeta|q,\nu_{n}^{(g)}\rangle\langle p,\nu_{n}^{(g)}|\zeta'\rangle\langle\zeta'|j,\nu_{n'}^{(g)}\rangle.\label{eq:chi_in_vibronic_basis}
\end{eqnarray}

Our goal is to write $S_{PP'}(T)$ for arbitrary bandwidth in a similar
style, so that in the broadband limit, we can identify it as a linear
combinations of elements of $\chi(T)$, hence proving Eqs. (\ref{eq:SE})--(\ref{eq:GSB})
in the article. We start by rewriting Eqs. (\ref{eq:0})--(\ref{eq:PPP'P'})
in the vibronic bases:

\begin{eqnarray}
|\Psi_{PP'}(s)\rangle & = & -\sum_{iq}(\boldsymbol{\mu}_{gi}\cdot\boldsymbol{e}_{P'})(\boldsymbol{\mu}_{qg}\cdot\boldsymbol{e}_{P})\sum_{\zeta n'}\langle i,\nu_{n'}^{(g)}|\zeta\rangle\langle\zeta|q,\nu_{n}^{(g)}\rangle e^{-i\omega_{\zeta}(s-t_{1})}\nonumber \\
 &  & \times\left(\sum_{m}\langle\nu_{m}^{(f)}|\nu_{n'}^{(g)}\rangle\tilde{\epsilon}_{P'}(\omega_{fm,\zeta})\tilde{\epsilon}_{P}(\omega_{\zeta,gn})|f\rangle+\tilde{\epsilon}_{P'}(\omega_{\zeta,gn'})\tilde{\epsilon}_{P}(\omega_{\zeta,gn})|g\rangle\right)|\nu_{n}^{(g)}\rangle,\label{eq:PP'_vibronic}\\
|\Psi_{PP}(s)\rangle & = & -\sum_{iq}(\boldsymbol{\mu}_{ig}\boldsymbol{\cdot}\boldsymbol{e}_{P})(\boldsymbol{\mu}_{qg}\cdot\boldsymbol{e}_{P})\sum_{\zeta n'}\langle i,\nu_{n'}^{(g)}|\zeta\rangle\langle\zeta|q,\nu_{n}^{(g)}\rangle e^{-i\omega_{gn'}(s-t_{1})}\nonumber \\
 &  & \times\frac{1}{2}\tilde{\epsilon}_{P}(\omega_{gn',\zeta})\tilde{\epsilon}_{P}(\omega_{\zeta,gn})\left(1-\mbox{Erf}\left(\frac{i\sigma((\omega_{L}+\omega_{gn',\zeta})+(\omega_{L}-\omega_{\zeta,gn}))}{2}\right)\right)|g\rangle|\nu_{n'}^{(g)}\rangle,\label{eq:PP_vibronic}\\
|\Psi_{P'P'}(s)\rangle & = & -\sum_{jp}(\boldsymbol{\mu}_{ig}\cdot\boldsymbol{e}_{P'})(\boldsymbol{\mu}_{pg}\cdot\boldsymbol{e}_{P'})\sum_{\zeta n'}\langle j,\nu_{n'}^{(g)}|\zeta\rangle\langle\zeta|p,\nu_{n}^{(g)}\rangle e^{-i\omega_{gn'}(s-t_{2})}\nonumber \\
 &  & \times\frac{1}{2}\tilde{\epsilon}_{P}(\omega_{gn',\zeta})\tilde{\epsilon}_{P}(\omega_{\zeta,gn})\left(1-\mbox{Erf}\left(\frac{i\sigma((\omega_{L}+\omega_{gn',\zeta})+(\omega_{L}-\omega_{\zeta,gn}))}{2}\right)\right)|g\rangle|\nu_{n'}^{(g)}\rangle,\label{eq:P'P'_vibronic}\\
|\Psi_{PPP'P'}(s)\rangle & = & \sum_{ijqp}(\boldsymbol{\mu}_{gj}\boldsymbol{\cdot}\boldsymbol{e}_{P})(\boldsymbol{\mu}_{pg}\cdot\boldsymbol{e}_{P})(\boldsymbol{\mu}_{gi}\boldsymbol{\cdot}\boldsymbol{e}_{P})(\boldsymbol{\mu}_{qg}\cdot\boldsymbol{e}{}_{P})\nonumber \\
 &  & \times\sum_{\zeta'\zeta nn'n''}\langle j,\nu_{n''}^{(g)}|\zeta'\rangle\langle\zeta'|p,\nu_{n'}^{(g)}\rangle\langle i,\nu_{n'}^{(g)}|\zeta\rangle\langle\zeta|q,\nu_{n}^{(g)}\rangle e^{-i\omega_{gn'}(s-t_{1})}\nonumber \\
 &  & \times\frac{1}{4}\tilde{\epsilon}_{P'}(\omega_{gn'',\zeta'})\tilde{\epsilon}_{P'}(\omega_{\zeta',gn'})\tilde{\epsilon}_{P}(\omega_{gn',\zeta})\tilde{\epsilon}_{P}(\omega_{\zeta,gn})\nonumber \\
 &  & \times\left(1-\mbox{Erf}\left(\frac{i\sigma((\omega_{L}+\omega_{gn'',\zeta})+(\omega_{L}-\omega_{\zeta',gn'})}{2}\right)\right)\nonumber \\
 &  & \times\left(1-\mbox{Erf}\left(\frac{i\sigma((\omega_{L}+\omega_{gn',\zeta})+(\omega_{L}-\omega_{\zeta,gn})}{2}\right)\right)|g\rangle|\nu_{n''}^{(g)}\rangle.\label{eq:PPP'P'_vibronic}
\end{eqnarray}
 where the Erf functions appear due to pulse overlap. Eqs. (\ref{eq:SE_overlap})---(\ref{eq:GSB_overlap})
together with Eqs. (\ref{eq:PP'_vibronic})---(\ref{eq:PPP'P'_vibronic})
yield:

\begin{eqnarray}
S_{SE}(T) & = & \sum_{ijqp}(\boldsymbol{\mu}_{gi}\boldsymbol{\cdot}\boldsymbol{e}_{P'})(\boldsymbol{\mu}_{qg}\cdot\boldsymbol{e}_{P})(\boldsymbol{\mu}_{gp}\cdot\boldsymbol{e}_{P})(\boldsymbol{\mu}_{jg}\cdot\boldsymbol{e}_{P'})\sum_{\zeta\zeta'nn'}\tilde{\epsilon}_{P'}(\omega_{gn',\zeta})\tilde{\epsilon}_{P}(\omega_{\zeta,gn})\tilde{\epsilon}_{P}(\omega_{gn,\zeta'})\tilde{\epsilon}_{P'}(\omega_{\zeta',gn'})\nonumber \\
 &  & \times p_{n}e^{-i(\omega_{\zeta}-\omega_{\zeta'})T}\langle i,\nu_{n'}^{(g)}|\zeta\rangle\langle\zeta|q,\nu_{n}^{(g)}\rangle\langle p,\nu_{n}^{(g)}|\zeta'\rangle\langle\zeta'|j,\nu_{n'}^{(g)}\rangle,\label{eq:S_SE_vibronic}\\
S_{ESA}(T) & = & -\sum_{ijqp}(\boldsymbol{\mu}_{fi}\boldsymbol{\cdot}\boldsymbol{e}_{P'})(\boldsymbol{\mu}_{qg}\cdot\boldsymbol{e}_{P})(\boldsymbol{\mu}_{gp}\boldsymbol{\cdot}\boldsymbol{e}_{P})(\boldsymbol{\mu}_{jf}\cdot\boldsymbol{e}_{P'})\nonumber \\
 &  & \times\sum_{\zeta\zeta'nn'n''m}\langle\nu_{n''}^{(g)}|\nu_{m}^{(f)}\rangle\langle\nu_{m}^{(f)}|\nu_{n'}^{(g)}\rangle\tilde{\epsilon}_{P'}(\omega_{fm,\zeta})\tilde{\epsilon}_{P}(\omega_{\zeta,gn})\tilde{\epsilon}_{P}(\omega_{gn,\zeta'})\tilde{\epsilon}_{P'}(\omega_{\zeta',fm})\nonumber \\
 &  & \times p_{n}e^{-i(\omega_{\zeta}-\omega_{\zeta'})T}\langle i,\nu_{n'}^{(g)}|\zeta\rangle\langle\zeta|q,\nu_{n}^{(g)}\rangle\langle p,\nu_{n}^{(g)}|\zeta'\rangle\langle\zeta'|j,\nu_{n''}^{(g)}\rangle,\label{eq:S_ESA_vibronic}\\
S_{GSB}(T) & = & 2\Re\sum_{ijqp}(\boldsymbol{\mu}_{gi}\cdot\boldsymbol{e}_{P'})(\boldsymbol{\mu}_{qg}\cdot\boldsymbol{e}_{P'})(\boldsymbol{\mu}_{gp}\cdot\boldsymbol{e}_{P})(\boldsymbol{\mu}_{jg}\boldsymbol{\cdot}\boldsymbol{e}_{P})\left(\frac{1}{4}\right)\sum_{\zeta\zeta nn'}\tilde{\epsilon}_{P'}(\omega_{gn',\zeta})\tilde{\epsilon}_{P'}(\omega_{\zeta,gn})\tilde{\epsilon}_{P}(\omega_{gn,\zeta'})\tilde{\epsilon}_{P}(\omega_{\zeta',gn'})\nonumber \\
 &  & \times\Bigg\{ p_{n}\langle i,\nu_{n'}^{(g)}|\zeta\rangle\langle\zeta|q,\nu_{n}^{(g)}\rangle\langle p,\nu_{n}^{(g)}|\zeta'\rangle\langle\zeta'|j,\nu_{n'}^{(g)}\rangle e^{-i\omega_{gn,gn'}T}\nonumber \\
 &  & \times\left(1-\mbox{Erf}\left(\frac{i\sigma((-\omega_{gn',\zeta}-\omega_{L})+(\omega_{\zeta,gn}-\omega_{L}))}{2}\right)\right)\left(1-\mbox{Erf}\left(\frac{i\sigma((-\omega_{gn,\zeta'}-\omega_{L})+(\omega_{\zeta',gn'}-\omega_{L}))}{2}\right)^{*}\right)\nonumber \\
 &  & +p_{n}\langle i,\nu_{n'}^{(g)}|\zeta\rangle\langle\zeta|q,\nu_{n}^{(g)}\rangle\left(\langle p,\nu_{n}^{(g)}|\zeta'\rangle\langle\zeta'|j,\nu_{n'}^{(g)}\rangle e^{i\omega_{gn',gn}T}\right)^{*}\nonumber \\
 &  & \times\left(1-\mbox{Erf}\left(\frac{i\sigma((-\omega_{gn',\zeta}-\omega_{L})+(\omega_{\zeta,gn}-\omega_{L}))}{2}\right)\right)\left(1-\mbox{Erf}\left(\frac{i\sigma((-\omega_{gn,\zeta'}-\omega_{L})+(\omega_{\zeta',gn'}-\omega_{L}))}{2}\right)\right)\Bigg\}.\label{eq:S_GSB_vibronic}
\end{eqnarray}
 The expressions above can be intuitively understood in terms of the
double-sided Feynman diagrams in Fig. S2. The expression for GSB consists
of a sum of terms corresponding to two types of Feynman pathways which
are different in general.

\begin{figure}
\begin{centering}
\includegraphics[scale=0.4]{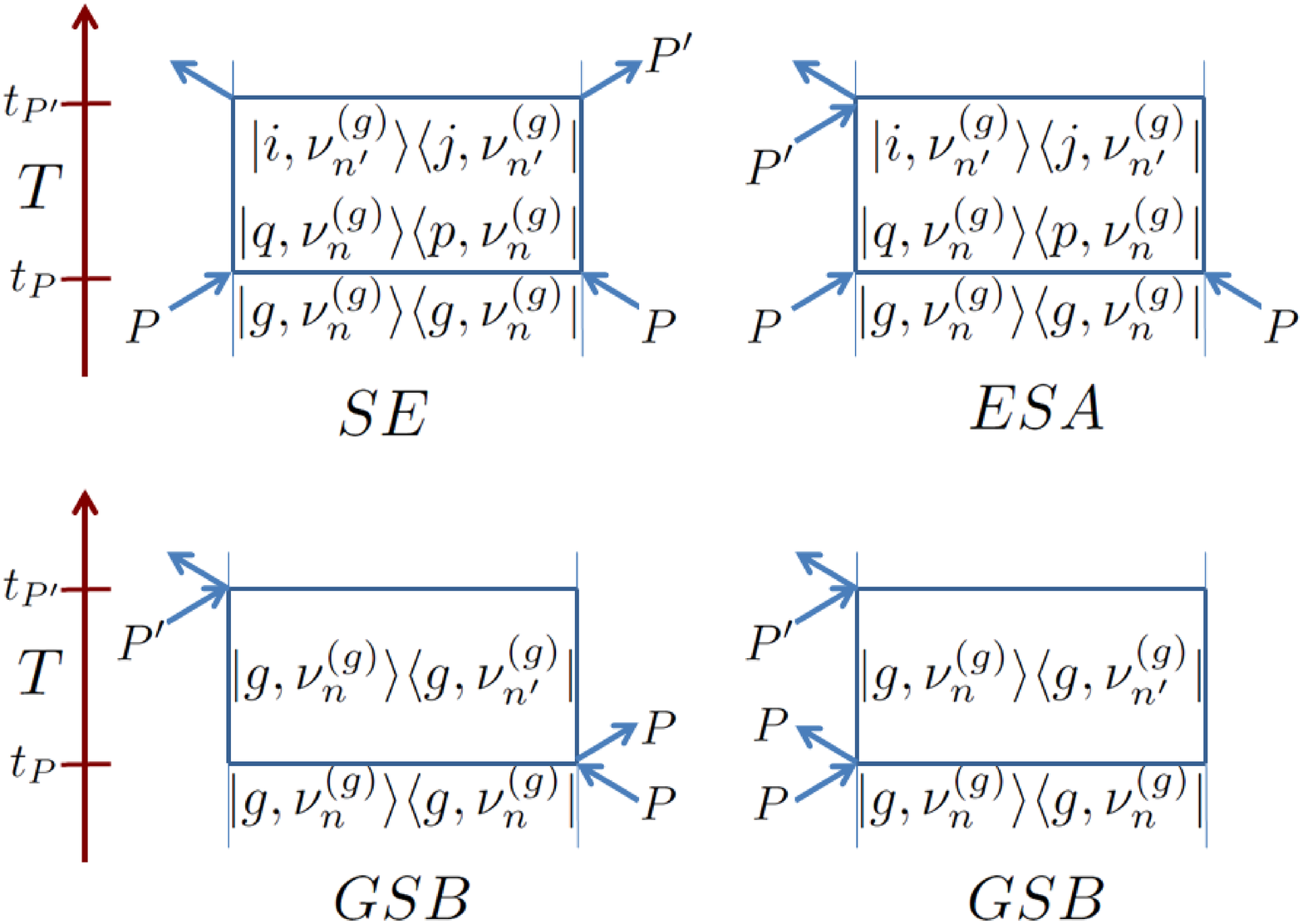}
\par\end{centering}

\caption{Double-sided Feynman diagrams for general PP' signal.}
\end{figure}

In the broadband limit where $\tilde{\epsilon}_{P}(\omega)=\tilde{\epsilon}_{P'}(\omega)=\lambda$,
many sums above collapse through resolutions of the identity, and
we straightforwardly recover the expressions in the article,

\begin{eqnarray}
S_{SE}^{bb}(T) & = & \lambda^{4}\sum_{ijpq}(\boldsymbol{\mu}_{gi}\cdot\boldsymbol{e}_{P'})(\boldsymbol{\mu}_{qg}\cdot\mbox{\ensuremath{\boldsymbol{e}}}_{P})\label{eq:SE}\\
 &  & \times(\boldsymbol{\mu}_{gp}\cdot\mbox{\ensuremath{\boldsymbol{e}}}_{P})(\boldsymbol{\mu}_{jg}\cdot\boldsymbol{e}_{P'})\chi_{ijqp}(T),\nonumber \\
S_{ESA}^{bb}(T) & = & -\lambda^{4}\sum_{ijpq}(\boldsymbol{\mu}_{fi}\cdot\boldsymbol{e}_{P'})(\boldsymbol{\mu}_{qg}\cdot\mbox{\ensuremath{\boldsymbol{e}}}_{P})\label{eq:ESA}\\
 &  & \times(\boldsymbol{\mu}_{gp}\cdot\mbox{\ensuremath{\boldsymbol{e}}}_{P})(\boldsymbol{\mu}_{jf}\cdot\boldsymbol{e}_{P'})\chi_{ijqp}(T),\nonumber \\
S_{GSB}^{bb}(T) & = & \lambda^{4}\sum_{ip}(\boldsymbol{\mu}_{gp}\cdot\mbox{\ensuremath{\boldsymbol{e}}}_{P})(\boldsymbol{\mu}_{pg}\cdot\mbox{\ensuremath{\boldsymbol{e}}}_{P})\label{eq:GSB}\\
 &  & \times(\boldsymbol{\mu}_{gi}\cdot\boldsymbol{e}_{P'})(\boldsymbol{\mu}_{ig}\cdot\boldsymbol{e}_{P'}).\nonumber
\end{eqnarray}
EIn this limit, as highlighted by the $T$-independent form of Eq.
(\ref{eq:GSB}), the two types of GSB pathways yield the same stationary
background to the signal (caused by copies of the initial stationary
wavepackets in the ground electronic surface).

In the practical case where the pulses are broad, but not infinitely
sharp in time, we can expand, $S_{PP'}(T)=S_{PP'}^{bb}(T)+S_{PP'}^{(1)}(T)$
where $S_{PP'}^{(1)}(T)=S_{GSB}^{(1)}(T)$ corresponds to corrections
of $O(\sigma)$, which originate from the Erf functions in the GSB
signal:

\begin{eqnarray*}
S_{GSB}^{(1)}(T) & = & -\frac{\lambda^{4}}{2}\sum_{ijqp}(\boldsymbol{\mu}_{gi}\cdot\boldsymbol{e}_{P'})(\boldsymbol{\mu}_{qg}\cdot\boldsymbol{e}_{P'})(\boldsymbol{\mu}_{gp}\cdot\boldsymbol{e}_{P})(\boldsymbol{\mu}_{jg}\boldsymbol{\cdot}\boldsymbol{e}_{P})\\
 &  & \times\Re\Bigg\{\sum_{\zeta\zeta nn'}p_{n}\langle i,\nu_{n'}^{(g)}|\zeta\rangle\langle\zeta|q,\nu_{n}^{(g)}\rangle\langle p,\nu_{n}^{(g)}|\zeta'\rangle\langle\zeta'|j,\nu_{n'}^{(g)}\rangle e^{-i\omega_{gn,gn'}T}\\
 &  & \times\frac{i((-\omega_{gn',\zeta}-\omega_{L})+(\omega_{\zeta,gn}-\omega_{L}))}{\sqrt{\pi}}+\frac{-i((-\omega_{gn,\zeta'}-\omega_{L})+(\omega_{\zeta',gn'}-\omega_{L}))}{\sqrt{\pi}}\\
 &  & +\sum_{\zeta\zeta nn'}p_{n}\langle i,\nu_{n'}^{(g)}|\zeta\rangle\langle\zeta|q,\nu_{n}^{(g)}\rangle\left(\langle p,\nu_{n}^{(g)}|\zeta'\rangle\langle\zeta'|j,\nu_{n'}^{(g)}\rangle e^{i\omega_{gn',gn}T}\right)^{*}\\
 &  & +\frac{i((-\omega_{gn',\zeta}-\omega_{L})+(\omega_{\zeta,gn}-\omega_{L}))}{\sqrt{\pi}}+\frac{i((-\omega_{gn,\zeta'}-\omega_{L})+(\omega_{\zeta',gn'}-\omega_{L}))}{\sqrt{\pi}}\Bigg\}\sigma.
\end{eqnarray*}

SE and ESA processes only contribute to corrections of $O(\sigma^{2})$
via the Gaussian spectral profile of the pulses.

\section{Requirement of broadband pump P}

Although the conclusions of the premilinary example in the article
hold even in the case of narrowband P', we also require broad bandwidth
for P for two reasons:
\begin{enumerate}
\item \emph{Non-stationary GSB contributions.} Eqs. (\ref{eq:PP}) and (\ref{eq:PPP'P'})
show that in the limit of broadband P, this pulse promotes a wavepacket
to the excited states and immediately back down to $|g\rangle$, yielding
a wavefunction $|\Psi_{PP}(t)\rangle$ that is proportional to the
original $|\Psi_{0}(t_{P})\rangle$ before any pulse (also see Eq.
\ref{eq:GSB}). In this limit, as emphasized in the previous section,
$S_{GSB}(T)$ is a constant background as a function of $T$, giving
the opportunity to identify $S_{PP'}(T)$ as a probe for singly-excited
state dynamics. Under a narrowband P, this no longer holds, as shown
by Eq. (\ref{eq:S_GSB_vibronic}), which depends on $T$ in general.
In this case, $|\Psi_{PP}(t)\rangle$ will be a non-stationary wavepacket
in the ground electronic surface, which will manifest as time-evolving
overlaps both in $\langle\Psi_{PP}(s)|g\rangle\langle g|\Psi_{P'P'}(s)\rangle$
and in $\langle\Psi_{0}(s)|g\rangle\langle g|\Psi_{PPP'P'}(s)\rangle$
(see Eq. (\ref{eq:GSB_overlap})).
\item \emph{Consistency with QPT.} As mentioned in the article, the initial
states prepared under a broadband P are of the form $\rho(0)=|q\rangle\langle p|\otimes\rho_{B}(0)$,
where $\rho_{B}(0)=\sum_{n}p_{n}|\nu_{n}^{(g)}\rangle\langle\nu_{n}^{(g)}|$
is the initial thermal ensemble of vibrations in the ground electronic
surface for all $|q\rangle\langle p|$. Under a narrowband P, it is
not possible to prepare initial tensor product states between the
system and a fixed bath $\rho_{B}$, so $S_{PP'}(T)$ can no longer
be written in terms of elements of a single $\chi(T)$, and the equation
$\rho(T)=\chi(T)\rho(0)$ loses its meaning.
\end{enumerate}

\section{Polaron transformation}

Here, we summarize the essential features of the polaron transformation
used in the arguments of the article. We closely follow the works
of Silbey, Harris, and coworkers \cite{silbey-harris,harris:1069,jang_polaron}.
Consider the approximation where the diabatic potential energy surfaces
are given by harmonic wells along each nuclear coordinate, $V_{10}(\boldsymbol{R})=E_{10}+\sum_{n}\frac{m_{n}\omega_{n}^{2}R_{n}^{2}}{2}+\sqrt{2m_{n}\omega_{n}^{3}}g_{10,n}R_{n}$,
and $V_{01}(\boldsymbol{R})$ has the same form except for the substitution
$10\to01$ in the subscripts, whereas $V_{g}(\boldsymbol{R})$ and
$V_{f}(\boldsymbol{R})$ have arbitrary shapes. Here, $\omega_{n}$
denotes the $n$-th mode frequency, whereas the displacements $g_{10,n}$
denote linear couplings of the electronic system to the nuclear bath.
Define the harmonic oscillator creation and anhilation operators in
the usual way $b_{n}^{\underbar{\ensuremath{\dagger}}}=\sqrt{\frac{m_{n}}{2}}x_{n}\mp\frac{1}{\sqrt{2m_{n}\omega_{n}}}\frac{\partial}{\partial x_{n}}$,
and also the generator $G=\sum_{n}(b_{n}^{\dagger}-b_{n})(g_{10,n}|10\rangle\langle10|+g_{01,n}|01\rangle\langle01|)$
such that $U=e^{G}$ corresponds to a unitary transformation of the
full-polaron transformation \cite{jang_polaron}. It follows that
$\tilde{H}(\boldsymbol{R})\equiv e^{G}H_{0}(\boldsymbol{R})e^{-G}=\tilde{H}_{0}(\boldsymbol{R})+\tilde{H}_{1}(\boldsymbol{R})$,
where $\tilde{H}_{0}(\boldsymbol{R})=T_{N}+\tilde{H}_{el}(\boldsymbol{R})$
is our new zeroth-order Hamiltonian, and $\tilde{H}_{1}(\boldsymbol{R})$
is the perturbation term, whenever it is small compared to $\tilde{H}_{0}(\boldsymbol{R})$.
To make a connection with the previous notation, we explicitly write
$\tilde{H}_{el}=\sum_{mn}\tilde{V}_{mn}|mn\rangle\langle mn|+\tilde{J}(|10\rangle\langle01|+|01\rangle\langle10|)$,
where,

\begin{eqnarray*}
\tilde{V}_{10} & = & E_{10}-\sum_{n}\omega_{n}g_{10,n}^{2}+\sum_{n}\frac{m_{n}\omega_{n}^{2}R_{n}^{2}}{2},\\
\tilde{V}_{01} & = & E_{01}-\sum_{n}\omega_{n}g_{01,n}^{2}+\sum_{n}\frac{m_{n}\omega_{n}^{2}R_{n}^{2}}{2},\\
\tilde{J} & = & J\langle w\rangle,\\
w & = & \mbox{exp}\left(\sum_{n}(g_{10,n}-g_{01,n})(b_{n}^{\dagger}-b_{n})\right),\\
\langle w\rangle & = & \mbox{Tr}(w\rho_{B}(0))\\
 & = & \mbox{exp}\left(-\sum_{n}\mbox{coth}\frac{\beta\omega_{n}(g_{nD}-g_{nA})^{2}}{2}\right).
\end{eqnarray*}
 The expressions $\beta$, $\rho_{B}(0)=\prod_{n}\sum_{r}\frac{\mbox{exp}\left(-\beta\omega_{n}\left(r+\frac{1}{2}\right)\right)|\omega_{n},r\rangle\langle\omega_{n},r|}{Z_{n}(\beta)}$,
$Z_{n}(\beta)=\frac{1}{2\mbox{sinh}(\beta\omega_{n}/2)}$, $|\omega_{n},r\rangle$
label the inverse temperature, the initial thermal ensemble of vibrations,
the partition function of the $n$-th oscillator, and the $r$-th
eigenstate of the $n$-th harmonic oscillator, respectively. $\tilde{J}$
can be interpreted as a renormalized site-site coupling due to phonon-dressing.
Furthermore, $\tilde{H}_{1}(\boldsymbol{R})=J(w-\langle w\rangle)|10\rangle\langle01|+\mbox{h.c.}$
Jang advises to consider the smallness of the quantity $J\sqrt{1-w^{2}}$
as the figure of merit for the validity of perturbation theory, and
hence for the usefulness of the polaron transformation. Cao and coworkers
note that the accuracy of the polaron transformation is guaranteed
only in the scenario of fast bath decorrelation compared to the other
relevant timescales \cite{cao_accuracy}.

If the dynamics of all the degrees of freedom are governed by $\tilde{H}_{0}(\boldsymbol{R})$
alone, the electronic system is effectively uncoupled from the nuclear
bath. The diagonalization of $\tilde{H}_{0}(0)$ yields polaronic
states $\{|g\rangle,|\tilde{\alpha}\rangle,|\tilde{\beta}\rangle,|f\rangle\}$
which satisfy $\chi_{ijqp}(T)=\delta_{iq}\delta_{jp}e^{-i\omega_{qp}T}$.
As can be easily checked, Eqs. (\ref{eq:SE}---\ref{eq:GSB}) are
invariant under change of basis. Hence, if $\tilde{H}_{1}(\boldsymbol{R})$
can be guaranteed to be a small perturbation for $\tilde{H}_{0}(\boldsymbol{R})$,
to zeroth-order in $\tilde{H}_{1}(\boldsymbol{R})$, the coherent
electronic oscillations in $S_{PP'}^{bb}(T)$ correspond to electronic
coherences in the polaronic basis.

The steps above have been outlined for the full-polaron transformation,
but the conclusion can be easily seen to hold whenever the total Hamiltonian
can be repartitioned into a large contribution and a small system-bath
coupling. Examples include the variational polaron transformation
\cite{silbey-harris,harris:1069}, which interpolates between weak
and strong coupling between the original system and bath, as well
transformations which include anharmonicities in the diabatic potential
energy surfaces (quadratic coupling between the original system and
bath, \cite{munn:2439}).

\section{Details of the numerical simulations in the article}

We have performed computational simulations %
\footnote{Details of the computational methodology will be presented elsewhere.%
} for absorption spectra, PP' signal, and rephasing 2D-ES for a monomer,
a dimer which exhibits electronic coherent oscillations, and an incoherent
dimer. For their Hamiltonians, we choose harmonic diabatic surfaces
parametrized by $V_{mn}(x,y)=E_{mn}+\frac{\omega_{mn,x}^{2}(x-\Delta_{mn,x})^{2}}{2}+\frac{\omega_{mn,y}^{2}(y-\Delta_{mn,y})^{2}}{2}$,
where $x$ and $y$ are scaled nuclear coordinates, $E_{mn}$ are
site energies, $\omega_{mn,x(y)}$ are oscillator frequencies and
$\Delta_{mn,x(y)}$ are electron-nuclear couplings \cite{forster,cina_fleming,cinakilinhumble,cinabiggs1,cinabiggs2}.
The parameters for the calculations are listed in Table 1. We assumed
that $V_{f}(x,y)=V_{10}(x,y)+V_{01}(x,y)$, and that the carrier frequency
of all the pulses is $\omega_{p}=\omega_{L}$. The dimers are such
that the dipoles are oriented 90 degrees from each other, and the
ratio between their norms is 1:3.

Fig. S1 shows inhomogeneously and rotationally averaged absorption
spectra (solid red) for the examples used in the numerical simulations
of the article. The pulse profiles (dotted black lines) which roughly
cover the strongest vibronic transitions with similar amplitude, giving
a qualitative idea of what broadband means in practice. In these examples,
$\mbox{FWHM}=2\sqrt{2ln2}\sigma=10,\,18.7,$ and $12.5$ fs, for the
monomer, coherent dimer, and incoherent dimer cases, respectively.

\begin{center}
\begin{tabular}{cccc}
\hline
\multicolumn{4}{c}{TABLE 1. Parameters of simulations}\tabularnewline
\hline
 & M  & CD  & ID\tabularnewline
$E_{00}/\mbox{cm}^{-1}$  & $0$  & $0$  & $0$\tabularnewline
$\langle E_{10}-\omega_{L}\rangle/\mbox{cm}^{-1}$  & $-125$  & $-300$  & $-200$\tabularnewline
$\langle E_{01}-\omega_{L}\rangle/\mbox{cm}^{-1}$  & ---  & $-200$  & $-200$\tabularnewline
$J/\mbox{cm}^{-1}$  & ---  & $100$  & $10$\tabularnewline
$\omega_{00,x}=\omega_{00,y}/\mbox{cm}^{-1}$  & $100$  & $100$  & $100$\tabularnewline
$\omega_{10,x}=\omega_{01,x}/\mbox{cm}^{-1}$  & $200$  & $200$  & $200$\tabularnewline
$\omega_{10,y}=\omega_{01,y}/\mbox{cm}^{-1}$  & $150$  & $150$  & $150$\tabularnewline
$\Delta_{00,x}/\mbox{cm}^{1/2}$  & $0$  & $0$  & $0$\tabularnewline
$\Delta_{00,y}/\mbox{cm}^{1/2}$  & $0$  & $0$  & $0$\tabularnewline
$\Delta_{10,x}/\mbox{cm}^{1/2}$  & $100$  & $50$  & $100$\tabularnewline
$\Delta_{10,y}/\mbox{cm}^{1/2}$  & $0$  & $0$  & $0$\tabularnewline
$\Delta_{01,x}/\mbox{cm}^{1/2}$  & $0$  & $0$  & $0$\tabularnewline
$\Delta_{01,y}/\mbox{cm}^{1/2}$  & $100$  & $50$  & $100$\tabularnewline
$\mbox{FWHM=}2\sqrt{2\mbox{ln}2}\sigma/\mbox{fs}$  & $10$  & $18.7$  & $12.5$\tabularnewline
\hline
\end{tabular}
\par\end{center}

\begin{figure}
\begin{centering}
\includegraphics[scale=0.45]{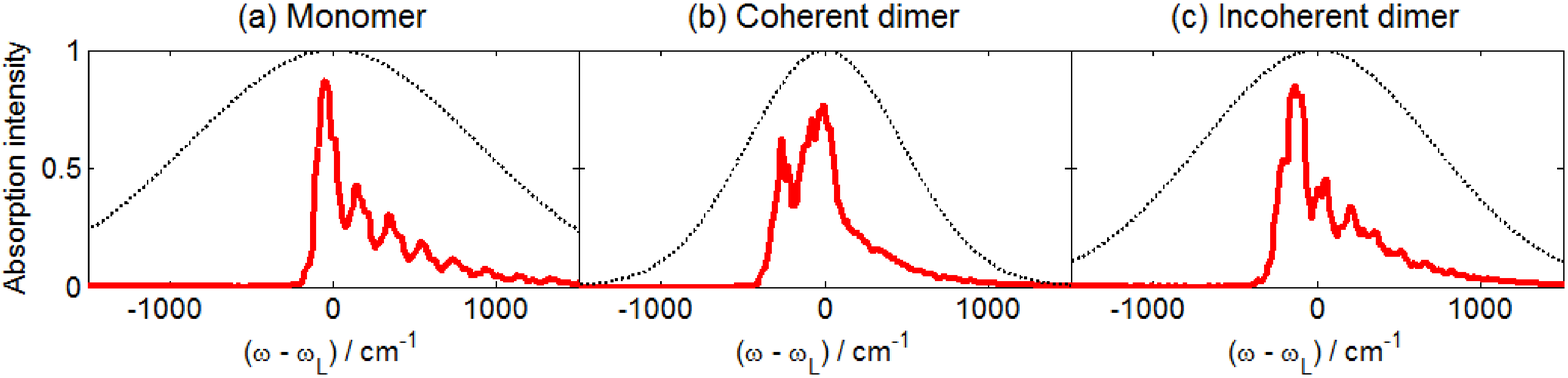}
\par\end{centering}

\caption{Inhomogeneously broadened absorption spectra (solid red) with pulse
spectral profiles $|\tilde{\epsilon}_{p}(\omega)|^{2}$ on top (dotted
black).}
\end{figure}

\bibliographystyle{unsrt}

\begin{thebibliography}{References}
\bibitem{mohseni} M.~Mohseni, P.~Rebentrost, S.~Lloyd, and A.~Aspuru-Guzik,
\newblock Environment-assisted quantum walks in photosynthetic energy
transfer. \newblock {\em J. Chem. Phys.}, 129:174106, 2008.

\bibitem{plenio} M.~B. Plenio and S.~F. Huelga, \newblock {\em
New J. Phys.}, 10:113019, 2008.

\bibitem{engelfleming} G.~S. Engel, T.~R. Calhoun, E.~L. Read,
T.~K. Ahn, T.~Mancal, Y.~C. Cheng, R.~E. Blankenship, and G.~R.
Fleming, \newblock {\em {Nature}}, {446}:{782--786}, {2007}.

\bibitem{scholes} E.~Collini, C.~Y. Wong, K.~E. Wilk, P.~M.~G.
Curmi, P.~Brumer, and G.~D. Scholes, \newblock {\em {Nature}},
{463}:{644--U69}, {2010}.

\bibitem{engelchicago} G.~Panitchayangkoon, D.~Hayes, K.~A. Fransted,
J.~R. Caram, E.~Harel, J.~Wen, R.~E. Blankenship, and G.~S. Engel,
\newblock {\em Proc. Natl. Acad. Sci. USA}, 107(29):12766--12770,
2010.

\bibitem{egorova} D.~Egorova. \newblock {\em Chem. Phys.}, 347(1-3):166
-- 176, 2008.

\bibitem{kaufmann_group} N.~Christensson, F.~Milota, J.~Hauer,
J.~Sperling, O.~Bixner, A.~Nemeth, and H.~F. Kauffmann, \newblock
{\em J. Phys. Chem. B}, 115(18):5383--5391, 2011.

\bibitem{2012arXiv1201.6325C} N.~{Christensson}, H.~F. {Kauffmann},
T.~{Pullerits}, and T.~{Mancal}, \newblock {\em arXiv:1201.6325
{[}quant-ph{]}}.

\bibitem{biggs_cina_anharmonic} J.~D. Biggs and J.~A. Cina, \newblock
{\em J. Phys. Chem. A}, 116(7):1683--1693, 2012.

\bibitem{chin_plenio} A.~W. {Chin}, J.~{Prior}, R.~{Rosenbach},
F.~{Caycedo-Soler}, S.~F. {Huelga}, and M.~B. {Plenio}, \newblock
{\em arXiv:1203.0776 {[}quant-ph{]}}.

\bibitem{moran_vibrations} J.~M. Womick and A.~M. Moran, \newblock
{\em J. Phys. Chem. B}, 115(6):1347--1356, 2011.

\bibitem{Shim2012649} S.~Shim, P.~Rebentrost, S.~Valleau, and
A.~Aspuru-Guzik, \newblock {\em Biophys. J.}, 102(3):649 -- 660,
2012.

\bibitem{cheng} Y.~C. Cheng and G.~R. Fleming, \newblock {\em
{J. Phys. Chem. A}}, {112}:{4254--4260}, {2008}.

\bibitem{turner} D.~B. Turner, K.~E. Wilk, P.~M.~G. Curmi, and
G.~D. Scholes, \newblock {\em J. Phys. Chem. Lett.}, 2(15):1904--1911,
2011.

\bibitem{Panitchayangkoon27122011} G.~Panitchayangkoon, D.~V. Voronine,
D.~Abramavicius, J.~R. Caram, N.~H.~C. Lewis, S.~Mukamel, and
G.~S. Engel, \newblock {\em Proc. Nat. Acad. Sci. USA}, 108(52):20908--20912,
2011.

\bibitem{cinabiggs1} J.~D. Biggs and J.~A. Cina, \newblock {\em
J. Chem. Phys.}, 131:224101, 2009.

\bibitem{yuen-aspuru} J.~Yuen-Zhou and A.~Aspuru-Guzik, \newblock
{\em J. Chem. Phys.}, 134(13):134505, 2011.

\bibitem{yuenzhou} J.~{Yuen-Zhou}, J.~J.~{Krich}, M.~{Mohseni},
and A.~{Aspuru-Guzik}, \newblock {\em Proc. Nat. Acad. Sci.
USA}, 108(43):17615,2011.

\bibitem{forster} T.~Forster, in \newblock {\em Modern Quantum
Chemistry}, volume~3, pages 93--137, \newblock Academic Press Inc.,
New York, 1965.

\bibitem{mukamel} S.~Mukamel, \newblock {\em Principles of Nonlinear
Optical Spectroscopy}, \newblock Oxford University Press, 1995.

\bibitem{Supplemental material} Supplemental material for this article.

\bibitem{minhaengbook} M.~Cho, \newblock {\em Two Dimensional
Optical Spectroscopy}, \newblock CRC Press, 2009.

\bibitem{tannor} D.~J. Tannor, \newblock {\em Introduction to
Quantum Mechanics: A Time Dependent Approach}, \newblock University
Science Books, 2007.

\bibitem{cina_fleming} J.~A. Cina and G.~R. Fleming, \newblock
{\em J. Phys. Chem. A}, 108(51):11196--11208, 2004.

\bibitem{lidar} D.~A. Lidar, I.~L. Chuang, and K.~B. Whaley, \newblock
{\em Phys. Rev. Lett.}, 81:2594--2597, Sep 1998.

\bibitem{yan_mukamel} Y.~J. Yan and S.~Mukamel, \newblock {\em
Phys. Rev. A}, 41:6485--6504, Jun 1990.

\bibitem{silbey-harris} R.~Silbey and R.~A. Harris, \newblock
{\em J. Chem. Phys.}, 80(6):2615--2617, 1984.\end{thebibliography}

\begin{thebibliography}{10}

\bibitem{mukamel}
S.~Mukamel,
\newblock {\em Principles of Nonlinear Optical Spectroscopy},
\newblock Oxford University Press, 1995.

\bibitem{mukamel_2d}
S.~Mukamel,
\newblock {\em Ann. Rev. Phys. Chem.}, 51(1):691--729, 2000.

\bibitem{minhaengbook}
M.~Cho,
\newblock {\em Two Dimensional Optical Spectroscopy},
\newblock CRC Press, 2009.

\bibitem{egorova}
D.~Egorova,
\newblock {\em Chem. Phys.}, 347(1-3):166 -- 176, 2008.

\bibitem{mukamel_adv}
S.~Mukamel, C.~Ciordas-Ciurdariu, and V.~Khidekel,
\newblock {\em Advances in Chemical Physics}, pages 345--372.
\newblock John Wiley and Sons, Inc., 2007.

\bibitem{cina_fleming}
J.~A. Cina and G.~R. Fleming,
\newblock {\em J. Phys. Chem. A}, 108(51):11196--11208,
  2004.


\bibitem{cinabiggs1}
J.~D. Biggs and J.~A. Cina,
\newblock {\em J. Chem. Phys.}, 131:224101, 2009.

\bibitem{cinabiggs2}
J.~D. Biggs and J.~A. Cina,
\newblock {\em J. Chem. Phys.}, 131:224302, 2009.

\bibitem{biggs_cina_anharmonic}
J.~D. Biggs and J.~A. Cina,
\newblock {\em J. Phys. Chem. A}, 116(7):1683--1693, 2012.

\bibitem{cao_wilson}
Jianshu Cao and Kent~R. Wilson,
\newblock {\em J. Chem. Phys.}, 106(12):5062--5072, 1997.

\bibitem{soo}
S.Y. Lee, in
\newblock {\em Femtosecond Chemistry}, pages 273--298.
\newblock Wiley-VCH Verlag GmbH, 2008.

\bibitem{silbey-harris}
R.~Silbey and R.~A. Harris,
\newblock {\em J. Chem. Phys.}, 80(6):2615--2617, 1984.

\bibitem{harris:1069}
R.~A. Harris and R.~Silbey,
\newblock {\em J. Chem. Phys.}, 83(3):1069--1074, 1985.

\bibitem{jang_polaron}
S.~Jang,
\newblock {\em J. Chem. Phys.}, 131(16):164101, 2009.

\bibitem{cao_accuracy}
C.~{Kong Lee}, J.~{Moix}, and J.~{Cao},
\newblock {\em arXiv:1201.2436 [quant-ph]}.

\bibitem{munn:2439}
R.~W. Munn and R.~Silbey,
\newblock {\em J. Chem. Phys.}, 68(5):2439--2450, 1978.


\bibitem{forster} T.~Forster, in \newblock {\em Modern Quantum
Chemistry}, volume~3, pages 93--137, \newblock Academic Press Inc.,
New York, 1965.

\bibitem{cinakilinhumble}
J.~A. Cina, D.~S. Kilin, and T.~S. Humble,
\newblock {\em J. Chem. Phys.}, 118:46--61, 2003.

\end{thebibliography}

\end{document}